\documentstyle[11pt]{article}
\newcommand{\be}{\begin{equation}}
\newcommand{\ee}{\end{equation}}

\title{SOME REMARKS ON TIME, UNCERTAINTY, AND
SPIN}
\author{Robert Carroll\\University of Illinois\\ email: 
rcarroll@math.uiuc.edu}

\date{March, 1999}

\begin{document}
\bibliographystyle{plain}
\maketitle
\begin{abstract}  
Some observations are made about energy-time uncertainty and
spin in the context of trajectories as in Faraggi-Matone or Floyd.
\end{abstract}

\section{INTRODUCTION}
\renewcommand{\theequation}{1.\arabic{equation}}
\setcounter{equation}{0}

The background here is properly the Bohmian or trajectory approach
to quantum mechanics (QM) (cf. \cite{bm,ha} for technical details
and \cite{bg,cm,hb,pd} for perspectives, philosophy, etc.).  In a
seminal paper \cite{fa} Faraggi and Matone (FM) develop a 1-D
trajectory theory based on a deep equivalence principle which seems to
provide the proper foundational structure for such theories (cf. also
\cite{fh,fi}).   For example quantization is a direct consequence of
the equivalence principle and there is a nontrivial action even
for bound states. 
In particular in \cite{fa} one avoids a flaw in the Bohm
theory based on the erroneous assumption that particle velocity
$\dot{q}$ is the same as $p/m=\partial_qS_0=S_0'$ where $S_0=W$ is
Hamilton's characteristic function or reduced action ($S=S_0-Et$).
The correct version here is $p=\partial_qW=m\partial_{\tau}q$ where
$\tau-\tau_0=m\int_{q_0}^q(dx/\partial_xW)$ represents a time
concept developed by Floyd (cf. \cite{fb,fc,fd,fe,fn,
fr}) in studies of trajectory representations and microstates.
In particular one can work with $t\sim\partial_EW$ to write
$t-t_0=\partial_E\int_{q_0}^qW'dx$ and arrive at 
$m\dot{q}=m(dt/dq)^{-1}=
m/W'_E=W'/(1-\partial_EQ)$ where $Q$ is the quantum potential
$Q=(\hbar^2/4m)\{W,q\}$ (Schwartzian derivative - details below).
Given even alone this important variation from the traditional Bohm
theory (and more generally, given \cite{fa}) many philosophical
discussions (some quite recent) regarding trajectory representations
should be drastically modified.  We will mostly avoid philosophy
in this note and simply make a few comments about matters related
to this use of time.  One notes in \cite{fa} also that a formula
$p=m_Q\dot{q}$ can be obtained via the use of a quantum mass
$m_Q=m(1-\partial_EQ)$.  The quantum potential $Q$ is regarded
here as the particles reaction to an external potential $V$ and no
pilot-wave philosophy is needed.  
We will consider stationary states but allow $E$ to
vary continuously (so discrete eigenvalues $E_n$ are 
not indicated although some arguments could be adjusted to
include them).

\section{BACKGROUND}
\renewcommand{\theequation}{2.\arabic{equation}}
\setcounter{equation}{0}

We sketch briefly some background.  First one starts with 
$i\hbar\psi_t=-(\hbar^2/2m)\psi_{qq}+V\psi$ for 
stationary states $\psi=\psi(q)
exp(-iEt/\hbar)$ so $i\hbar\psi_t=E\psi$ and $-(\hbar^2/2m)\psi''
+V\psi=E\psi$ ($'\sim\partial_q$).  ``Classically" one takes then
$\psi=Rexp(iW/\hbar)$ where $S=S_0-Et$ and $S_0=W$ to 
arrive at
\be
\frac{1}{2m}(W')^2+V+Q-E=0;\,\,(R^2W')'=0
\label{1}
\ee
where $(\bullet)\,\,Q=-\hbar^2R''/2mR$ is the quantum potential.
In \cite{fa} this whole procedure is changed and the matter is
developed in the context of a general equivalence principle leading
to a quantum stationary Hamilton-Jacobi equation (QSHJE)
\be
\frac{1}{2m}(W')^2+{\cal W}(q)+Q(q)=0
\label{2}
\ee
which is actually an identity.  The individual terms are (here
$\{f,q\}=(f'''/f')-(3/2)(f''/f')^2$ is the Schwartzian derivative)
\be
{\cal W}(q)=-\frac{\hbar^2}{4m}\left\{exp\left(\frac{2iW}{\hbar}
\right),q\right\};\,\,Q(q)=\frac{\hbar^2}{4m}\{W,q\}
\label{3}
\ee
This $Q$ is the general quantum potential which in the special context
of (\ref{1}) is $Q=-(\hbar^2/2m)(R''/R)$ as in $(\bullet)$ (note
$R^2W'=c$ from (\ref{1}) will produce the Schwartzian derivative
$\{W,q\}$ as in (\ref{3})).  Further one can identify $(\bullet\bullet)\,\,
{\cal W}(q)=V-E$ for which we refer to \cite{fa}.  Thus 
given $V$ one can determine $W$ via $(\bullet\bullet)$ or via 
(\ref{2}) with ${\cal W}=V-E$.  Note that (\ref{2}) is a third order
differential equation for $W$ and its solutions will lead to microstates
\`a la Floyd.
\\[3mm]\indent
Now in Floyd's work \cite{fb} it was apparently first observed
that Bohm's assumption $p=W'=m\dot{q}$ (for particle velocity
$\dot{q}$) is not generally valid and the correct version is
(cf.  \cite{fa})
\be
m(1-\partial_EQ)\dot{q}=W'\equiv m_Q\dot{q}=W'\equiv
m\partial_{\tau}q=W';
\label{4}
\ee
$$m_Q=m(1-\partial_EQ);\,\,\tau-\tau_0=m\int_{q_0}^q\frac{dx}
{W'}$$
Then one has, using (\ref{1}) and Floyd's effective or modified
potential $U=V+Q$,
\be
t-t_0=\partial_E\int_{q_0}^qW'dx=\left(\frac{m}{2}\right)^{1/2}
\int_{q_0}^q\frac{(1-\partial_EQ)dx}{\sqrt{E-U}}
\label{5}
\ee
and $d\tau/dt=1/(1-\partial_EQ)$.  Thus $t$ is explicitly a function
of $E$ and we want to expand upon this aspect of the theory.
It is important to note that general solutions of
the Schr\"odinger equation above should be taken in the form
${\bf (2A)}\,\,\psi=(W')^{-1}[Aexp(-iW/\hbar)+Bexp(iW/\hbar)]$
and $p\sim\partial_qW=W'$ is the generic form for $p$ 
corresponding to momentum in a quantum mechanical Hamiltonian
$(1/2m)p^2+V\sim (1/2m)(-i\hbar\partial_q)^2+V$.  Thus $p=W'$
corresponds to $p\sim -i\hbar\partial_q$ and this is the quantum
mechanical meaning for $p$; it will not correspond in general to
mechanical momentum $m\dot{q}$ for particle motion.

\section{GENERAL COMMENTS IN 1-D}
\renewcommand{\theequation}{3.\arabic{equation}}
\setcounter{equation}{0}

First one sees that Hamilton-Jacobi (HJ) procedures involve
$t\sim\partial_EW$ and we will modify an argument in \cite{fz}
in order to give further insight into the relation $t=t(E)$.  We think
of a general stationary state situation with $S\sim W-Et=W(q,E)-Et$
so that $\partial S/\partial t=-E$ and $t=t(E)$.  Setting ${\cal S}=
-S$ with $\partial{\cal S}/\partial t=E$ we can write then 
$(\clubsuit)\,\,W=Et-{\cal S}=t{\cal S}_t-{\cal S}$ in Legendre form.
Now given $W=W(E,q)$, with $q$ fixed, one has $\partial_EW=t
+Et_W-{\cal S}_tt_W=t$ so $(\spadesuit)\,\,{\cal S}=EW_E-W$
gives the dual Legendre relation.  Consequently the constructions
in \cite{fa} for example automatically entail the Legendre transformation
relations $(\clubsuit)$ and $(\spadesuit)$ involving ${\cal S}=-S$ and
$W$.
\\[3mm]\indent
Now one comes to the energy-time uncertainty ``principle" which should
be mentioned because of situations involving energy 
dependent time for example (cf. \cite{aa,bk,da,ea,fs,se} for various
approaches - we make no attempt to be
complete or exhaustive here).  First, in a perhaps simple minded
spirit, let us recall that microstates are compatible with the 
Schr\"odinger representation by wave functions $\psi$ and hence
one will automatically have a connection of the trajectory representation
with Hilbert space ideas of observables and probability 
(more on this below).  In the 
Hilbert space context the uncertainty $\Delta q\Delta p\geq\hbar/2$ is
a trivial consequence of operator inequalities and we take it to mean
nothing more nor less ($\Delta p$ for example represents a standard
deviation $<\psi|(\hat{p}-<\hat{p}>)^2|\psi>$ where $\hat{p}\sim
-i\hbar\partial_q$).  In this spirit nothing need be said about 
measurement and we will not broach the subject in any way except
to say that sometimes for a trajectory we will think e.g. of physical
increments $\delta q\sim q-q_0$ and $\delta p\sim p-p_0$.  Thus
we will try to maintain a distinction between $\delta q$ and $\Delta q$
for example and we do not require that $\delta q$ be measured, only
that it be a natural mathematical concept.  After all $\Delta q$ above
is also only a natural mathematical concept without any a priori
connection to measurement.  The idea of attaching physical meaning
to $\Delta q$ via measurement seems no stranger than thinking
of $\delta q$ as a meaningful possibly measurable physical
quantity.
As for energy-time uncertainty we remark first that if one departs
by $\epsilon$ from a correspondence between observables and 
self-adjoint operators then an $\epsilon$ approximate inequality
${\bf (ET)}\,\,\Delta E\Delta t\geq\hbar/2$ can be proved in a
Hilbert space context (see e.g. \cite{fs} for a detailed discussion). 
There are also various crude physical derivations 
based on $\delta q\sim (p/m)\delta t$  where $p$ is the 
physical momentum and
subsequently, for $\delta E\sim (p/m)\delta p$ when e.g. $E\sim
(p^2/2)+V(q)$, one often writes ${\bf (EET)}\,\,
\delta p\delta q\sim(p/m)(m/p)
\delta E\delta t=\delta E\delta t
\geq\hbar/2$ based on a $q-p$ uncertainty with
$\delta q\sim\Delta q$ etc. (displacement version). 
This would be fine if $p=m\dot{q}$ but we have seen that $p$ has 
an unambiguous quantum mechanical meaning as in Section 2 and
the argument involving $\delta q=(p/m)\delta t$ is generally
not valid. 
A more convincing argument can be made
via use of Ehrenfest type relations ($\hat{H}\sim E$)
\be
\frac{d<\hat{Q}>}{dt}=\frac{1}{\hbar}<i[\hat{H},\hat{Q}]>;\,\,
\delta E\delta Q\geq\frac{\hbar}{2}\left|\frac{d}{dt}<\hat{Q}>
\right|
\label{6}
\ee
and an argument that the time $\delta t$ for a change $\delta Q=
\delta <\hat{Q}>$ should be $\delta t=\delta Q/|d<\hat{Q}>/dt|$,
leading to $
{\bf ({\cal E}{\cal E}{\cal T})}\,\,\delta E\delta t\geq\hbar/2$ 
(without intervention
of $p$, where however one has $d<q>/dt\sim (1/m)<p>$).
\\[3mm]\indent
Since the beginning step $\delta q\sim (p/m)\delta t$ is generally
wrong in the crude argument above
let us adjust this following (\ref{4}) to be $\delta q\sim
(p/m)\delta\tau\sim(p/m_Q)\delta t$ where $p=W'$ is the conjugate
momentum (QM momentum).
Then with $\delta E\sim
(p/m)\delta p$ one will arrive at $\delta p\delta q\sim\delta E
\delta \tau\sim\delta E\delta t/((1-Q_E)$ and consequently a
correct displacement (or perhaps trajectory)
version of {\bf (ET)} should be
\be
\delta E\delta \tau\geq\frac{\hbar}{2}\equiv\delta E\delta t
\geq (1-\partial_EQ)\frac{\hbar}{2}
\label{7}
\ee
Since in the trajectory picture
we are dealing with $t=t(E)$ via $t=\partial_EW$ (with
$E$ a continuous variable here) one will have $\delta t=W_{EE}
\delta E$ so (\ref{7}) requires a curious condition
$(\clubsuit\clubsuit)\,\,(\delta E)^2\geq [(1-Q_E)\hbar/2W_{EE}]$.
Thus apparently for any energy change compatible with the microstate
picture $(\clubsuit\clubsuit)$ must hold.  
This would seem to preclude any positive infinitesimal $\delta E$
unless $(1-Q_E)/W_{EE}\leq 0$ (with restrictions on any negative
$\delta E$).  One can perhaps envision here microstates as developed
by Floyd generated at energy $E_0$ with initial conditions
$W_0(E_0),\,W'_0(E_0),$ and $W''_0(E_0)$ (or $(q_0,\,\dot{q}_0,
\,\ddot{q}_0)(E_0)$) and then imagine $E$ to be changed while keeping
the initial conditions constant.  This would affect the time relations
on the trajectories and lead to a situation where the inequality
$(\clubsuit\clubsuit)$ could have meaning, but for general situations
$t=t(E)$ (\ref{7}) seems untenable, and hence we exclude 
energy-time uncertainty for completely determined microstates
(see below however).
Further we cannot suggest its
applicability in the operator form ${\bf (ETT)}\,\,\Delta E\Delta \tau
\geq \hbar/2$ since that would clash with {\bf (ET)} which has a
more or less substantial foundation.  Thus we argue that while
{\bf (ET)} may be acceptable its displacement version 
{\bf (EET)} is not, except perhaps in the averaged form
${\bf ({\cal E}{\cal E}{\cal T})}$.
\\[3mm]\indent
As for computation in (\ref{7}) for example
one notes that the equations in \cite{fe,fn,fr} for example
have to be put in ``canonical" form as in \cite{fa} and,
in computing $W_E$, one should
only differentiate terms which under
a transformation $E\to E'\not= E$ do not correspond to a M\"obius
transformation of $exp(2iW/\hbar)$ (i.e. one only differentiates
terms in which ${\cal W}$ is changed under a transformation $E\to E'
\ne E$). 
Regarding $1-Q_E$ one can use the relation $W'W'_E=m(1-Q_E)$ for
computation.   As for uncertainty
however my interpretation of some remarks of Floyd suggests the 
following approach.  First I would claim that uncertainty type inequalities
are incompatible with functional relations between the quantities
(e.g. $p=\partial_qW=p(q)$ or $t=t(E)$ via $W$).  Thus if
$W$ is completely known there is generally no room for uncertainty
since e.g. $\delta t\sim W_{EE}\delta E$ or with adjustment of
constants $\delta t=t-t_0=W_E$ completely specifies $\delta t$.  Note
that one of the themes in \cite{fa} involves replacing canonical
transformations between independent variables $(p,q)$ with
coordinate transformations $q\to\tilde{q}$ with $p=W_q(q)$ depending
on $q$.  Now we recall that the QSHJE is third order and one needs
three initial conditions $(q_0,\dot{q}_0,\ddot{q}_0)$ or $(W_0,W'_0,
W''_0)$ for example to determine a solution and fix the microstate
trajectories.  However the Copenhagen representation uses an 
insufficient set of initial conditions for microstates (and literally
precludes microstate knowledge).  The substitute for exact microstate
knowledge is then perhaps an uncertainty principle.
It would be interesting to see if the two
pictures interact and one could perhaps think of uncertainty
relations involving $\delta t$ and $\delta E$ 
as in (\ref{7}) for example
as constraints
for the microstate initial conditions.   However the microstates are
always compatible with the Schr\"odinger equation for any
initial conditions and hence lead to the same operator
conclusions in Hilbert space (such as 
{\bf (ET)} for example).
In any event one can continue to use the standard quantum mechanics,
knowing that a deliberate sacrifice of information has been made
in not specifying the background microstates (i.e. quantum
mechanics in Hilbert space is imprecise by construction, leading
naturally to a probabilistic theory etc.).  We refrain from 
any further attempts at ``philosophy" here.

\section{SPIN}
\renewcommand{\theequation}{4.\arabic{equation}}
\setcounter{equation}{0}

We follow here \cite{eb} with references to \cite{ec,rf,
rh,sf,ta} (a very incomplete list). 
First without discussion we write down some equations
from \cite{eb} and \cite{rh} ($\hbar$ is removed in \cite{eb} so
we reinsert it \`a la \cite{rh}).  Thus one thinks of $\psi=Rexp(iS/\hbar)$
with
\be
{\cal L}=-\rho\left[S_t+\frac{1}{2m}(\nabla S)^2+\frac
{\hbar^2}{8m}\left(\frac{\nabla\rho}{\rho}\right)^2+V\right]
\label{97}
\ee
($\rho=R^2$).
Thence one determines the equations for a Madelung fluid
\be
S_t+\frac{1}{2m}(\nabla S)^2+\frac{\hbar^2}{4m}\left[\frac{1}{2}
\left(\frac{\nabla\rho}{\rho}\right)^2-\frac{\Delta\rho}{\rho}\right]
+V=0;\,\,\rho_t+\nabla\cdot (\rho\nabla S/m)=0
\label{98}
\ee
(cf. also \cite{sf}).  The quantum potential is ($|\psi|=R=\rho^{1/2}$)
\be
\frac{\hbar^2}{4m}\left[\frac{1}{2}\left(\frac{\nabla\rho}{\rho}
\right)^2-\frac{\Delta\rho}{\rho}\right]=-\frac{\hbar^2}{2m}
\frac{\Delta |\psi|}{|\psi|}=Q
\label{99}
\ee
and this arises from the single nonclassical term $(\hbar^2/8m)
(\nabla\rho/\rho)^2$ in ${\cal L}$ of (\ref{97}).  The internal motion
or spin is related to the Zitterbewegung idea and $\vec{v}\not= 
\vec{p}/m$ in this context. 
Now one defines
\be
\vec{v}_B=\frac{\nabla S}{m};\,\,\vec{v}_S=\frac{\hbar\nabla\rho}
{2m\rho}
\label{9}
\ee
and we note that the equations are invariant under $\vec{J}=\rho
\vec{v}_B\to\vec{J}+\nabla\times\vec{b}$.  This leads to a spin
vector $\vec{s}$ with current velocity ${\bf (4A)}\,\,\vec{v}
=\vec{v}_B+\vec{v}_S\times\vec{s}=\vec{v}_{\parallel}+
\vec{v}_{\perp}$ and ${\bf (4B)}\,\,|\vec{s}|^2=1$ with $\vec{v}_S
\cdot\vec{s}=0$ and $\vec{v}_B\cdot (\vec{v}_S\times\vec{s})=0$.
One notes also that ${\bf (4C)}\,\,Q=-(m/2)\vec{v}_S^2-(\hbar/2)
\nabla\cdot\vec{v}_S$.  For stationary states $\psi=\psi(q)exp(-iEt/
\hbar)$ with $\psi(q)=Rexp(iW/\hbar)$ one has then
\be
\frac{1}{2m}(\nabla W)^2-\frac{\hbar^2}{2m}\frac{\Delta R}{R}
+V-E=0;\,\,\nabla\cdot (\rho\nabla W)=0
\label{10}
\ee
with $Q$ given in (\ref{99}).  
\\[3mm]\indent
We want to see now if we can relate the spin picture to the trajectory
representation.  Since we are dealing with the same basic
Schr\"odinger equation the only new feature is 3-D.  One can speak
of internal motion, spin, Zitterbewegung, etc. but once a current
velocity $\vec{v}$ appears as in {\bf (4A)} we are at least implicitly
making contact with the idea of particle motion and some comparison
between $\vec{v}$ and trajectory velocity $d\vec{q}/dt$ should be
possible and meaningful.  This point may be arguable but we assume
it momentarily at least.  Actually in \cite{eb} one explicitly deals
with $\vec{v}$ as a particle velocity so this should carry over to the
stationary state.  However the arguments in \cite{eb} about trajectory
representations do not take into account the work of FM or Floyd
involving microstates so we feel the conclusions in \cite{eb} should
be correspondingly adjusted (see below for more on this).  We will
show that the use of current velocity $\vec{v}$ as particle velocity
seems to be incorrect.
\\[3mm]\indent
Thus following the 1-D example of (\ref{4}) where $m(1-Q_E)\dot{q}
=W'$ we use the relation $t\sim\partial_EW$ again to suggest
($\vec{x}\sim\vec{q}$)
\be
t-t_0=\partial_E\int_{\vec{x}_0}^{\vec{x}}
\nabla W\cdot d\vec{x}\sim\nabla t=
\nabla W
\label{11}
\ee
Then from (\ref{10}) one has
\be
\frac{1}{n}\nabla W\cdot\nabla W_E+Q_E-1=0;\,\,\nabla\cdot
[\partial_E(\rho\nabla W)]=0
\label{12}
\ee
We can write then ${\bf (4D)}\,\,\partial_E(\rho\nabla W)=\nabla
\times\vec{\gamma}$ say and following $dt/dq=\partial_EW'=W'_E$ in
1-D, or $dq/dt=1/W'_E$, we suggest for (\ref{11}) the relation
${\bf (4E)}\,\,\partial t/\partial x_i=\partial W_E/\partial x_i\sim
\dot{x}_i=1/\partial_iW_E$.  In a similar manner (\ref{12}) leads to
\be
\nabla W\cdot\nabla t=m(1-Q_E);\,\,m\dot{x}_i=\frac{|\nabla W|^2}
{(1-Q_E)\partial_iW}
\label{13}
\ee
which implies
${\bf (4F)}\,\,m(1-Q_E)(d\vec{x}/dt)\cdot\nabla W=3|\nabla W|^2$.
\\[3mm]\indent
Now the constructions of \cite{eb} with $\vec{v}_B\cdot (\vec{v}_S
\times \vec{s})=0$ as in {\bf (4B)} give $\vec{v}\cdot\vec{v}_B=
|\vec{v}_B|^2$ where $\vec{v}_B=(1/m)\nabla W$.  If one could 
identify $\vec{v}$ with $d\vec{x}/dt$ there would result then from
{\bf (4F)} the formula ${\bf (4G)}\,\,(1-Q_E)=3m$ which is
unlikely at best.  Hence we conclude that the current velocity
$\vec{v}$ of \cite{eb} cannot be identified with particle velocity
and the conclusions there about trajectories are not correct.  Perhaps
the difficulty lies in the following observation.  Even though the quantum
potential $Q$ can be recovered from $\vec{v}_S$ as in {\bf (4C)}
nevertheless the full quantum potential is not used in constructing
$\vec{v}_S$ via (\ref{9}) as can be seen from (\ref{99}).  In the
trajectory picture from FM or Floyd the full quantum potential is used
in determining $d\vec{x}/dt$ as indicated in (\ref{13}) or {\bf (4F)}.
\\[3mm]\indent
A few additional comments should also be made about adapting the
development in \cite{eb} or \cite{rh} for stationary state situations.
Thus from (\ref{10}) we first extend $\vec{J}$ in the form $\vec{J}
=\rho\vec{v}_B\to \rho\vec{v}_B+\nabla\times (\rho\vec{c})$ and 
satisfy the equation $\nabla\cdot\vec{J}=0$ via ${\bf (4H)}\,\,
\vec{J}=\rho\vec{v}_B+\nabla\times (\rho\vec{c})=\
\nabla\times\vec{\phi}$ 
with $\vec{s}=(2m/\hbar)\vec{c}$ and $\vec{\phi}$ dependent on
$\vec{x}$.  Then (cf. (\ref{9})) $\nabla\cdot\vec{J}=0\sim \nabla
\cdot (\rho\vec{v}_B)=0$ and one can write 
$$\nabla\times (\rho\vec{c})=\nabla\rho\times\vec{c}+
\rho(\nabla\times\vec{c})=\rho\left[\vec{v}_S\times\vec{s}+
\frac{\hbar}{2m}\nabla\times\vec{s}\right]$$
with
\be
\vec{v}=\vec{v}_B+\vec{v}_S\times\vec{s};\,\,\vec{J}=\rho\vec{v}
+\vec{J}_0=\nabla\times\vec{\phi}=\vec{\eta};
\,\,\vec{J}_0=\frac{\hbar\rho}{2m}
\nabla\times\vec{s}
\label{14}
\ee
One can think of $\vec{J}_0$ as a ``pseudocurrent" added on to deal
with cases of a variable spin vector $\vec{s}(\vec{x})$ and $\vec{s}$
still is to satisfy {\bf (4B)}.  The term $\vec{\phi}$ is added here to
give a variable right side for $\vec{J}$.  Now in order to determine
if this produces a solvable configuration we note that the equations
{\bf (4B)} consist of three equations in three unknowns $s_i$
for $\vec{s}=(s_1,s_2,s_3)$ and the coefficients in these equations
depend on $\nabla W$ via $\vec{v}_B$ and on $\rho$ via $\vec{v}_S$
yielding $\vec{s}=\vec{s}(\rho,\nabla W)$.  We think of $\vec{\eta}$
as arbitrary for the moment and from ${\bf (4J)}\,\,\vec{v}=(1/\rho)
[\vec{\eta}-\vec{J}_0]$ one obtains via (\ref{98}) and (\ref{10})
\be
|\vec{v}|^2=|\vec{v}_B|^2+|\vec{v}_S|^2=\frac{2}{m}(E-V)
+\frac{\hbar^2}{2m^2}\frac{\Delta\rho}{\rho}=\left|\frac{\vec{\eta}}
{\rho}-\frac{\hbar}{2m}\nabla\times\vec{s}\right|^2
\label{15}
\ee
Thus we have two more equations, namely (\ref{15}) and {\bf (4J)}
in the form ${\bf (4K)}\,\,\vec{v}=(1/m)\nabla W+(\hbar/2m)(\nabla
\rho/\rho)=(\vec{\eta}/\rho)-(\hbar/2m)\nabla\times\vec{s}$.  If
we put $\vec{s}=\vec{s}(\rho,\nabla W)$ in these equations they
become two equations (\ref{15}) and {\bf (4K)} for $\rho$ and
$\nabla W$ in terms of an arbitrary $\vec{\eta}$.  Thus in principle
this configuration should be solvable
and some preliminary calculations
are promising.  As an example take $\vec{s}=\vec{i}$ and
$\nabla\rho=\rho_2\vec{j}$ so $curl\vec{s}=0$ and $\nabla\rho
\times\vec{s}=-\rho_2\vec{k}$.  Then $\vec{v}_B\perp
(\nabla\rho\times\vec{s})$ will imply $\nabla W=W_1\vec{i}
+W_2\vec{j}$ and one has ${\bf (4L)}\,\,\vec{\eta}/\rho=
(1/m)(W_1\vec{i}+W_2\vec{j})-(\hbar\rho/2m\rho)\vec{k}$.
Thus given $\vec{\eta}$ we must have
\be
\frac{\rho W_1}{m}=\eta_1;\,\,\frac{\rho W_2}{m}=\eta_2;\,\,
-\frac{\hbar\rho_2}{2m}=\eta_3
\label{16}
\ee
along with (\ref{15}) in the form
\be
\frac{2\rho^2}{m}(E-V)+\frac{\hbar^2\rho\rho_{22}}{2m^2}=
\frac{\rho^2}{m^2}(W_1^2+W_2^2)+\frac{\hbar^2\rho_2^2}{4m^2}
\label{17}
\ee
This can be satisfied if e.g. ${\bf (4M)}\,\,2(E-V)=W_1^2+W_2^2$
and $\rho\rho_{22}=(1/2)\rho_2^2$.  The latter equation is $\rho_{22}/
\rho_2=(1/2)(\rho_2/\rho)$ leading to $log(\rho_2/\rho^{1/2})=
f(x,z)$ and eventually $2\rho^{1/2}=\alpha(x,z)y +\beta(x,z)$
with $\alpha,\,\beta$ constant.  Evidently {\bf (4M)} requires also
$V=V(x,y)$.
\\[3mm]\indent
{\bf ACKNOWLEDGEMENT}$\,\,$  The author would like to
thank E. Floyd and M. Matone for valuable comments.


\newpage

\begin{thebibliography}{cc}

%
\bibitem{aa} Y. Aharonov and D. Bohm,
Phys. Rev., 122/5 (1961), 1649-1658; 134/6B (1964), 1417-1418
%
\bibitem{bg} D. Bohm and B. Hiley, The undivided universe, Routledge,
London, 1993
%
\bibitem{bm} D. Bohm,
Phys. Rev., 85 (1952), 166-179 and 180-193
%
\bibitem{bk} J. Briggs and J. Rost,
quant-ph 9902035
%
\bibitem{cm} J. Cushing,
Quantum mechanics, Univ. Chicago Press, 1994
%
\bibitem{da} V. Delgado and J. Muga,
Phys. Rev. A, 56 (1997), 3425-3435
%
\bibitem{ea} J. Eberly and L. Singh,
Phys. Rev. D, 7 (1973), 359-362
%
\bibitem{eb} S. Esposito,
quant-ph 9902019
%
\bibitem{ec} S. Esposito, 
Found. Phys., 28 (1998), 231-244; Phys. Lett. A, 225 (1997),203-209
%
\bibitem{fa} A. Faraggi and M. Matone, hep-th 9809127
%
\bibitem{fh} A. Faraggi and M. Matone, Phys. Rev. Lett., 78 (1997),
163-166
%
\bibitem{fi} A. Faraggi and M. Matone, hep-th 9705108 = Phys. Lett.
B, 450 (1999), 34; hep-th 9711028 = Phys. Lett. B, 437 (1998),
369; hep-th 9801033 = Phys. Lett. A, 249 (1998), 180; hep-th 
9809125 = Phys. Lett. B, 445 (1999), 77; and hep-th 9809126=
Phys. Lett. B, 445 (1999), 357
%
\bibitem{fz} B. Felsager,
Geometry, particles, and fields, Springer, 1998
%
\bibitem{fs} E. Fick and F. Engelmann,
Zeits. f\"ur Physik, 175 (1963), 271-282; 178 (1964), 551-562
%
\bibitem{fb} E. Floyd, Phys. Rev. D, 26 (1982), 1339-1347;
Phys. Rev. D, 29 (1984), 1842-1844; 
Phys. Rev. D, 25 (1982), 1547-1551
%
\bibitem{fc} E. Floyd, Phys. Lett. A, 214 (1996), 259-263
%
\bibitem{fd} E. Floyd,
Phys. Rev. D, 34 (1986), 3246-3249
%
\bibitem{fe} E. Floyd quant-ph 9707051; quant-ph 9708007;
quant-ph 9708026
%
\bibitem{fn} E. Floyd, Phys. Essays, 7 (1994), 135-145
%
\bibitem{fr} E. Floyd,
Ann. Fond. L. deBroglie, 20 (1995), 263-279
%
\bibitem{hb} B. Hiley and F. Peat, Quantum implications, Routledge,
London, 1987
%
\bibitem{ha} P. Holland, The quantum theory of motion, Cambridge
Univ. Press, 1997
%
\bibitem{mb} M. Morrison,
Understanding quantum mechanics, Prentice-Hall, 1990
%
\bibitem{pa} E. Papp,
The uncertainty principle and foundations of QM, Wiley, 1977,
pp. 29-50
%
\bibitem{pd} D. Park,
Classical dynamics and its quantum analogues, Springer, 1990
%
\bibitem{rf} M. Razavy,
Nuovo Cimento, 63 (1969), 271-308; Amer. Jour. Phys., 35
(1967),955-960
%
\bibitem{rh} E. Recami and G. Salesi,
Phys. Rev. D, 57 (1998), 98-105
%
\bibitem{sf} G. Salesi,
Mod. Phys. Lett. A, 11 (1996), 1815-1823
%
\bibitem{sg} G. Salesi and E. Recami,
Phys. Lett. A, 190 (1994), 137-143
%
\bibitem{se} F. Schwabl,
Quantum mechanics, Springer, 1995
%
\bibitem{ta} T. Takabayashi,
Prog. Theor. Phys., 69 (1983), 1323-1344
%



\end{thebibliography}
\end{document}